\documentclass[twocolumn,superscriptaddress,aps,preprintnumbers,amsmath,amssymb,nofootinbib
,a4paper
]{revtex4-2}

\usepackage{amsfonts,amsmath,amssymb,ascmac,bm,tensor}
\usepackage{fnpct} 
\usepackage{comment}
\usepackage{ifpdf}
\usepackage{slashed}
\usepackage{color}
\usepackage[mathscr]{eucal}
\usepackage{fourier}
\usepackage[utf8]{inputenc}
\usepackage{physics}
\usepackage{cancel}
\usepackage{subfig}
\usepackage{wasysym}
\usepackage{xcolor}

\usepackage{lipsum}
\ifpdf
  \usepackage{graphicx}     
  \usepackage[bookmarksopen,colorlinks=true,linkcolor=bblue,citecolor=bblue,urlcolor=ppink]{hyperref}
\else     
\fi

\usepackage{cancel}
\usepackage{ulem}
\usepackage{amsfonts}
\usepackage{amssymb}
\usepackage{graphicx}
\usepackage{amsmath,bm}
\usepackage{enumerate}
\usepackage{mathtools}
\usepackage{tikz} \usetikzlibrary{calc}
\usepackage{subfloat}
\usepackage{subfig}
\usepackage{xcolor}
\usepackage{float}
\usepackage{color}
\usepackage{here}

\usepackage{booktabs}
\usepackage{siunitx}
\usepackage[table]{xcolor}
\makeatletter
\renewcommand\@biblabel[1]{\textcolor{red}{[#1]}}

\let\old@bibitem\@bibitem
\renewcommand{\@bibitem}[2][]{%
  \old@bibitem[#1]{#2}%
  \color{blue}
}
\makeatother

\usepackage{adjustbox}
\usepackage{graphicx,tikz}
\usetikzlibrary{mindmap}

\usepackage{mathrsfs}
\usepackage{float,epsfig}
\usepackage{dcolumn}
\usepackage{pgfplots}
\usepackage{graphicx}
\usepackage{bm}
\usepackage{amsmath,amssymb,amsthm}

\usetikzlibrary{shadows}

\definecolor{red}{rgb}{1,0,0}
\definecolor{darkred}{rgb}{0.6,0,0}
\definecolor{darkgreen}{rgb}{0.992447,0.623778,0.034597}
\definecolor{ppink}{rgb}{1,0.4,0.4}
\definecolor{bblue}{rgb}{0.284602,0.317763,0.963947}
\definecolor{brown}{rgb}{0.5 ,0, 0.7}
\definecolor{blue2}{rgb}{0.2 ,0.2, 0.85}

\makeatletter
\newcommand\footnoteref[1]{\protected@xdef\@thefnmark{\ref{#1}}\@footnotemark}
\makeatother
 
\allowdisplaybreaks[1]

\definecolor{lime}{HTML}{A6CE39}
\newcommand{\orcidicon}{%
	\begin{tikzpicture}
	\draw[lime, fill=lime] (0,0)
	circle [radius=0.16]
	node[white] {{\fontfamily{qag}\selectfont \tiny ID}};
	\draw[white, fill=white] (-0.0625,0.095)
	circle [radius=0.007];
	\end{tikzpicture}   \hspace{-2mm}
}
\newcommand\orcidHasan{{\href{https://orcid.org/0000-0001-7408-0910}{\orcidicon}}}
\newcommand\orcidKarima{{\href{https://orcid.org/0000-0001-5419-8516}{\orcidicon}}}
\newcommand\orcidJamal{{\href{https://https://orcid.org/0000-0002-4463-4203}{\orcidicon}}}

\begin{document}


\title{
Critical Inter-Horizon Thermal Dynamics on the Lukewarm Reissner--Nordstr\"om--de Sitter Manifold
    }

\author{H. El Moumni \orcidHasan}	
\email{h.elmoumni@uiz.ac.ma (Corresponding author)}

\author{J. Khalloufi \orcidJamal\!\!}
\email{jamalkhalloufi@gmail.com}
\affiliation{\small LPTHE, Physics Department, Faculty of Sciences,  Ibnou Zohr University, Agadir, Morocco.}

\author{K.  Masmar \orcidKarima }
\email{k.masmar@gmail.com\\
(Authors are listed in alphabetical order.)}
\affiliation{\small LPTHE, Physics Department, Faculty of Sciences,  Ibnou Zohr University, Agadir, Morocco.}

\date{\today}

\begin{abstract}
We reinterpret the lukewarm sector of four-dimensional Reissner--Nordstr\"om--de Sitter black holes as the exact zero-dissipation thermal manifold of an effective two-horizon nonequilibrium system. In the fixed-charge sector, the inter-horizon thermal affinity controls the entropy production and vanishes precisely on the lukewarm branch. The corresponding linearized thermal mode is governed by an exact relaxation coefficient \(K_L(\rho)\), with \(\rho=r_+/r_c\), and changes stability at the critical ratio
\[
\rho_*=\frac{1+\sqrt{3}-\sqrt{2}\,3^{1/4}}{2}\approx 0.4354,
\]
where the relaxation time diverges as \(\tau\sim |\rho-\rho_*|^{-1}\). We then encode this critical structure in a minimal Bragg--Williams functional and an Onsager--Machlup action for the effective trajectories of the thermal mode. In this way, the lukewarm branch is promoted from a geometric equal-temperature locus to a critical inter-horizon thermal manifold.
\end{abstract}

\maketitle



\section{Introduction}

Black hole thermodynamics provides one of the sharpest windows into the interplay between gravitation, quantum theory, and statistical physics \cite{Bekenstein1973,Hawking1975}. In de Sitter space, however, the presence of multiple horizons with generically different temperatures obstructs a conventional global equilibrium description and makes multi-horizon black holes natural candidates for effective nonequilibrium thermodynamics \cite{GibbonsHawking1977}.

This problem has been explored from several complementary viewpoints. The lukewarm branch of Reissner--Nordstr\"om--de Sitter black holes, characterized by equal black hole and cosmological temperatures, is classical and well known \cite{Romans1992}. More broadly, de Sitter black hole thermodynamics has been studied through separated horizon first laws, effective temperatures, and global thermodynamic constructions \cite{UranoTomimatsuSaida2009,KubiznakSimovic2016,SimovicMann2019}. At the same time, recent developments have increasingly emphasized the intrinsically nonequilibrium nature of multi-horizon de Sitter systems.

The present work is neither a rederivation of the classical lukewarm branch nor a full semiclassical treatment of inter-horizon transport. Rather, it occupies an intermediate position between effective de Sitter thermodynamics and explicitly nonequilibrium descriptions, by reinterpreting the lukewarm branch as a critical thermal manifold of an effective two-horizon system. 
Within this framework, the inter-horizon thermal affinity supports a distinguished relaxation mode whose stability changes at a critical horizon ratio. The resulting critical structure is then encoded in a minimal Bragg--Williams functional and an Onsager--Machlup action \cite{Banerjee2010,OnsagerMachlup1953a,OnsagerMachlup1953b}, thereby promoting the lukewarm branch from a geometric equal-temperature locus to a critical inter-horizon thermal manifold.

\section{Lukewarm Geometry and Exact Thermal Manifold}
Consider the four-dimensional RN--dS geometry
\begin{equation}
ds^2=-f(r)\,dt^2+\frac{dr^2}{f(r)}+r^2 d\Omega_2^2,
\end{equation}
with
\begin{equation}
f(r)=1-\frac{2M}{r}+\frac{Q^2}{r^2}-\frac{\Lambda r^2}{3},
\qquad \Lambda>0.
\end{equation}
Let \(r_+\) and \(r_c\) denote the black hole and cosmological horizons, with \(0<r_+<r_c\). They satisfy
\begin{equation}
f(r_+)=0,
\qquad
f(r_c)=0.
\label{eq:horizon}
\end{equation}
The corresponding temperatures are
\begin{equation}
T_+=\frac{1}{4\pi}\left(\frac{1}{r_+}-\frac{Q^2}{r_+^3}-\Lambda r_+\right),
\label{eq:Tplus}
\end{equation}
and
\begin{equation}
T_c=\frac{1}{4\pi}\left(\Lambda r_c-\frac{1}{r_c}+\frac{Q^2}{r_c^3}\right).
\label{eq:Tc}
\end{equation}
The lukewarm condition reads
\begin{equation}
T_+=T_c.
\label{eq:lukewarm}
\end{equation}

\begin{figure}[!ht]
		\begin{center}
		\centering
			\begin{tabbing}
			\centering
			\hspace{9.3cm}\=\kill
			\includegraphics[scale=.5]{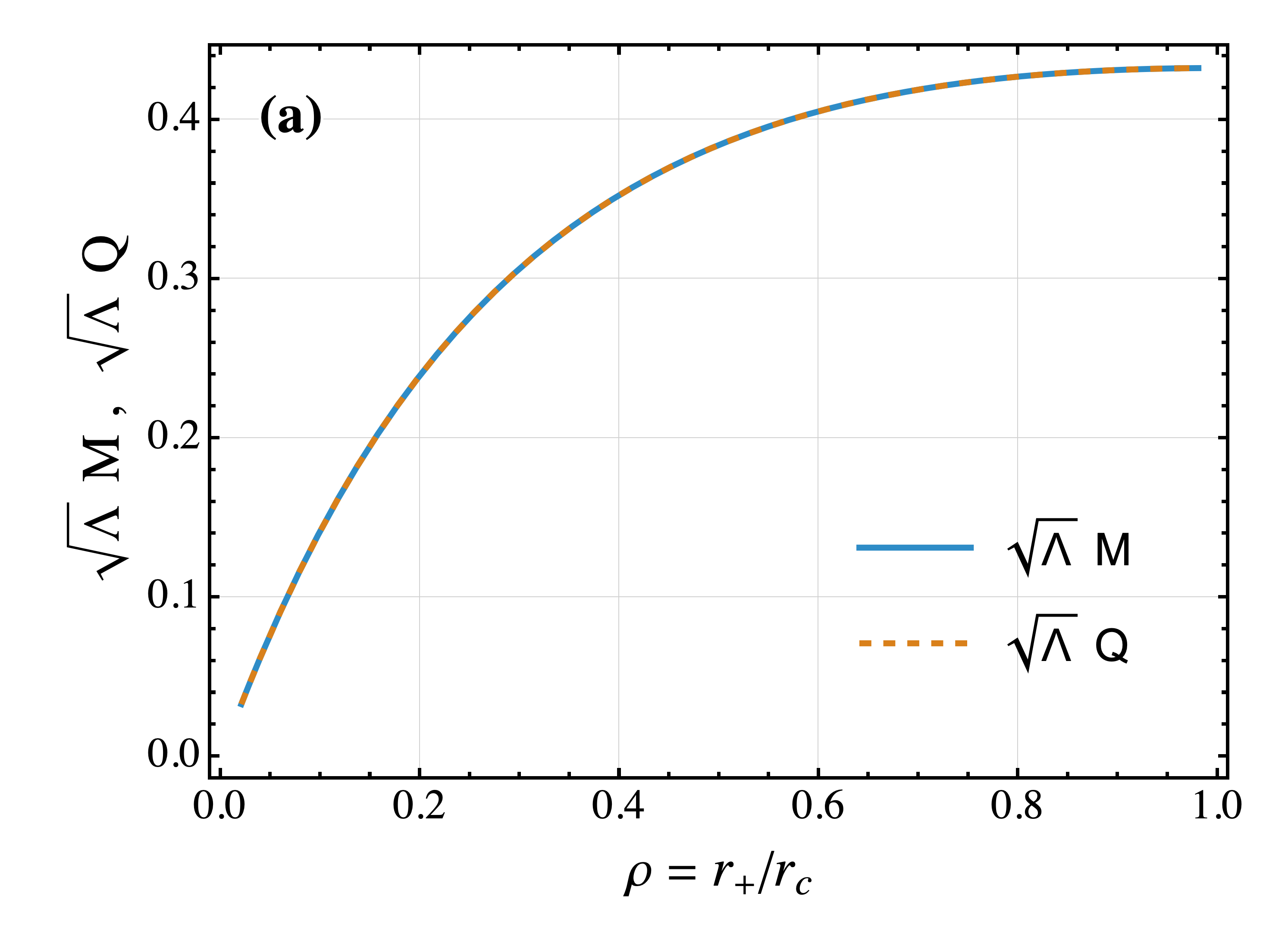} \\
			\includegraphics[scale=.5]{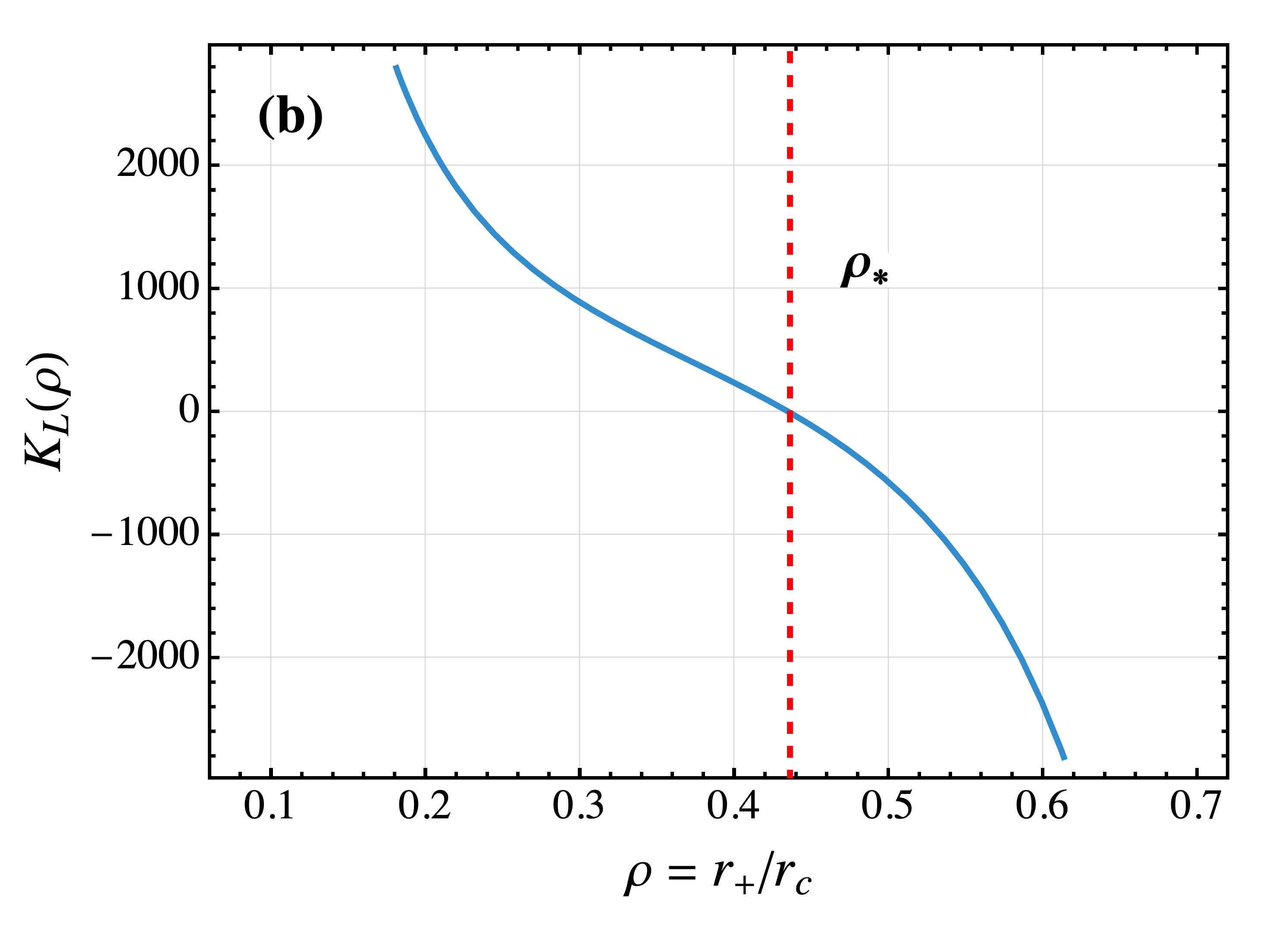} \\
		   \end{tabbing}

\caption{{\it \footnotesize  Lukewarm RN--dS black holes as a zero-dissipation critical manifold.
{\bf (a)} Exact lukewarm branch in dimensionless form, showing the coincidence \(\sqrt{\Lambda}M=\sqrt{\Lambda}Q\) along the manifold.
{\bf (b)} Relaxation coefficient \(K_L(\rho)\), whose sign change at \(\rho=\rho_*\) separates an attractive and a repulsive segment of the thermal fixed manifold.  }}\label{fig1}
\end{center}
\end{figure}

Multiplying the two horizon equations by $r_+$
 and $r_c$, respectively, and subtracting them, one obtains
\begin{equation}
Q^2
=
r_+r_c\left[
1-\frac{\Lambda}{3}(r_+^2+r_+r_c+r_c^2)
\right].
\label{eq:Qconstraint}
\end{equation}
Eliminating \(Q^2\) from Eq.~\eqref{eq:lukewarm} with Eq.~\eqref{eq:Qconstraint}, one obtains the exact factorization
\begin{equation}
(r_c-r_+)^2(r_c+r_+)\Big[\Lambda(r_++r_c)^2-3\Big]=0.
\end{equation}
Since \(r_c>r_+>0\), the physical branch is uniquely fixed by
\begin{equation}
\Lambda=\frac{3}{(r_++r_c)^2}.
\label{eq:Lambda}
\end{equation}
Substituting back yields
\begin{equation}
Q=\frac{r_+r_c}{r_++r_c},
\qquad
M=\frac{r_+r_c}{r_++r_c},
\label{eq:MQexact}
\end{equation}
so that
\begin{equation}
M=Q.
\end{equation}
The common lukewarm temperature becomes
\begin{equation}
T_L=T_+=T_c=\frac{r_c-r_+}{2\pi(r_++r_c)^2}.
\label{eq:TL}
\end{equation}
Thus the exact lukewarm manifold is
\begin{equation}
M=Q=\frac{r_+r_c}{r_++r_c},
\qquad
\Lambda=\frac{3}{(r_++r_c)^2}.
\label{eq:luke_manifold}
\end{equation}
By itself, this exact branch is not yet the main physical result; its significance lies in the nonequilibrium structure built upon it.
The exact lukewarm branch is shown in Fig.\ref{fig1}(a), where the dimensionless coincidence \(\sqrt{\Lambda}M=\sqrt{\Lambda}Q\) is made explicit along the manifold.

%

\section{Inter-Horizon Nonequilibrium Thermodynamics}
We now interpret the event and cosmological horizons as weakly coupled thermal subsystems in an effective quasistatic state-space dynamics at fixed \(Q\) and \(\Lambda\). This is a state-space dynamics, not a claim about an explicit time-dependent Einstein--Maxwell solution. Related thermodynamic treatments of de Sitter two-horizon systems have been explored in Refs.~\cite{UranoTomimatsuSaida2009,KubiznakSimovic2016,SimovicMann2019}. In this effective description, the two horizons are treated as thermodynamic subsystems coupled through an inter-horizon thermal exchange channel. The corresponding state-space dynamics is therefore organized by the same basic ingredients as in nonequilibrium thermodynamics: a thermal affinity, an effective current, and the associated entropy production. The purpose of this construction is not to provide a microscopic derivation of horizon transport, but to identify the minimal thermodynamic structure supported by the exact lukewarm manifold.

At fixed cosmological constant and in the fixed-charge sector, \(\delta Q=0\), the local first laws take the form
\begin{equation}
\delta M=T_+\,\delta S_+,
\qquad
\delta M=-T_c\,\delta S_c.
\label{eq:firstlaws_reduced}
\end{equation}
with
\begin{equation}
S_+=\pi r_+^2,
\qquad
S_c=\pi r_c^2,
\end{equation}
which are the entropies associated with the black hole and cosmological horizons, respectively.

Introducing an effective evolution parameter \(t\) along the quasistatic state-space dynamics, we define the inter-horizon thermal current by
\begin{equation}
J\equiv \dot M \equiv \frac{dM}{dt}.
\label{eq:Jdef}
\end{equation}
Here \(t\) is not the physical time coordinate of a dynamical Einstein--Maxwell spacetime, but an effective evolution parameter on the space of nearby thermodynamic configurations. 
We then obtain
\begin{equation}
\dot S_+=\frac{J}{T_+},
\qquad
\dot S_c=-\frac{J}{T_c}.
\end{equation}
Defining the total horizon entropy as \(S_{\rm tot}\equiv S_+ + S_c\), we arrive at
\begin{equation}
\dot S_{\rm tot}=J\left(\frac{1}{T_+}-\frac{1}{T_c}\right).
\label{eq:Sdot}
\end{equation}
This identifies the natural thermal affinity
\begin{equation}
X\equiv \beta_+-\beta_c=\frac{1}{T_+}-\frac{1}{T_c}.
\label{eq:X}
\end{equation}

As usual in nonequilibrium thermodynamics, we adopt the minimal linear-response constitutive law in the vicinity of a stationary thermal manifold. In this effective description, the inter-horizon current is therefore taken to be linear in the corresponding thermal affinity at leading order
\begin{equation}
J=L\,X+O(X^2),
\qquad
L>0,
\label{eq:J}
\end{equation}
so that
\begin{equation}
\dot S_{\rm tot}=L\,X^2+O(X^3),
\qquad
\dot S_{\rm tot}\ge 0 \;\text{to quadratic order}.
\label{eq:Sdot2}
\end{equation}
Hence the thermal fixed manifold is precisely \(X=0\), i.e. the lukewarm manifold. This result changes the status of the lukewarm branch. Rather than being viewed merely as a geometric equal-temperature locus, it emerges here as the stationary thermal manifold of the effective two-horizon system. In this sense, the lukewarm family is selected not only by exact horizon thermodynamics, but also by the vanishing of the leading nonequilibrium entropy production.

The departure from the lukewarm manifold can be characterized exactly. Eliminating \(Q^2\) via Eq.~\eqref{eq:Qconstraint}, one finds
\begin{equation}
4\pi(T_+-T_c)
=
\frac{(r_c-r_+)^2(r_c+r_+)}{3r_+^2r_c^2}
\Big[\Lambda(r_++r_c)^2-3\Big].
\label{eq:Tdiff}
\end{equation}
This motivates the exact dimensionless control parameter
\begin{equation}
\Xi\equiv \Lambda(r_++r_c)^2-3,
\label{eq:Xi}
\end{equation}
which vanishes exactly on the lukewarm manifold. 
The quantity \(\Xi\) geometrizes the departure from the lukewarm condition: it vanishes exactly on the thermal manifold and measures, in a fully geometric way, how far the two-horizon configuration lies from thermal balance. Near the manifold, the thermal affinity is proportional to \(\Xi\),
\begin{equation}
X=-A(r_+,r_c)\,\Xi+O(\Xi^2),
\label{eq:XXi}
\end{equation}
with
\begin{equation}
A(r_+,r_c)=\frac{\pi(r_++r_c)^5}{3r_+^2r_c^2}.
\end{equation}
Thus \(X\) and \(\Xi\) encode the same local departure from the lukewarm branch, but \(X\) is the more natural thermodynamic variable since it is directly conjugate to the entropy production.
\section{Critical Relaxation and Stability Threshold}

To determine the stability of the thermal fixed manifold, we define the horizon susceptibilities by
\begin{equation}
C_+=T_+\left(\frac{\partial S_+}{\partial T_+}\right)_{Q,\Lambda},
\qquad
C_c=-T_c\left(\frac{\partial S_c}{\partial T_c}\right)_{Q,\Lambda}.
\label{eq:susceptibilities}
\end{equation}
The minus sign in the cosmological sector reflects the opposite orientation already present in the local first law \(\delta M=-T_c\,\delta S_c\). With this convention, the quasistatic evolution equations take the form
\begin{equation}
C_+\,\dot T_+=J,\qquad C_c\,\dot T_c=J.
\label{eq:Tdot}
\end{equation}
Differentiating \(X=1/T_+-1/T_c\) then gives
\begin{equation}
\dot X
=
-J\left(
\frac{1}{C_+T_+^2}
-
\frac{1}{C_cT_c^2}
\right).
\label{eq:Xdot}
\end{equation}
Using Eq.~\eqref{eq:J}, we obtain the linear relaxation equation
\begin{equation}
\dot X=-L\,\mathcal K\,X,
\qquad
\mathcal K\equiv
\frac{1}{C_+T_+^2}
-
\frac{1}{C_cT_c^2}.
\label{eq:Xrelax}
\end{equation}

It is convenient to introduce
\begin{equation}
\rho=\frac{r_+}{r_c}\in(0,1).
\end{equation}
Restricting the general relaxation coefficient \(\mathcal K\) to the exact lukewarm manifold and substituting the corresponding horizon susceptibilities yields the closed expression
\begin{equation}
\mathcal K_L(\rho)
=
\frac{2\pi(1+\rho)^5}{\rho^3(1-\rho)^3}
\left(1-2\rho-2\rho^3+\rho^4\right).
\label{eq:KL}
\end{equation}
The coefficient \(\mathcal K_L(\rho)\) diverges at the endpoints \(\rho\to 0\) and \(\rho\to 1\), while its sign is controlled in the interior of the interval by the quartic factor \(1-2\rho-2\rho^3+\rho^4\).
 The behavior of the relaxation coefficient \(\mathcal K_L(\rho)\) is displayed in Fig.\ref{fig1}(b), where its sign change at \(\rho=\rho_*\) makes the transition from attractive to repulsive thermal behavior manifest.

Hence, the inter-horizon thermal mode satisfies
\begin{equation}
\dot X=-L\,\mathcal K_L(\rho)\,X.
\label{eq:mainmode}
\end{equation}
The critical ratio is determined by the unique physical root of the quartic factor,
\begin{equation}
\rho_*=
\frac{1+\sqrt3-\sqrt2\,3^{1/4}}{2}
\approx 0.4354205447.
\label{eq:rhostar}
\end{equation}
For \(0<\rho<\rho_*\), one has \(\mathcal K_L>0\), and \(X=0\) is linearly attractive. For \(\rho_*<\rho<1\), one has \(\mathcal K_L<0\), and the mode becomes linearly unstable. In the stable region, the relaxation time is defined by
\begin{equation}
\tau^{-1}=L\,\mathcal K_L(\rho).
\label{eq:tau}
\end{equation}
In the unstable region, \(|\tau|^{-1}=L\,|\mathcal K_L|\) defines the corresponding growth rate. Expanding near the critical ratio,
\begin{equation}
\mathcal K_L(\rho)=\mathcal K_L'(\rho_*)(\rho-\rho_*)+O((\rho-\rho_*)^2),
\end{equation}
with \(\mathcal K_L'(\rho_*)\neq 0\), one finds
\begin{equation}
\tau \sim |\rho-\rho_*|^{-1}.
\label{eq:slowing}
\end{equation}
The ratio \(\rho_*\) therefore defines a genuine critical marginal point of the inter-horizon thermal mode.

\begin{figure}[!ht]
		\begin{center}
		\centering
			\begin{tabbing}
			\centering
			\hspace{9.3cm}\=\kill
			\includegraphics[scale=.5]{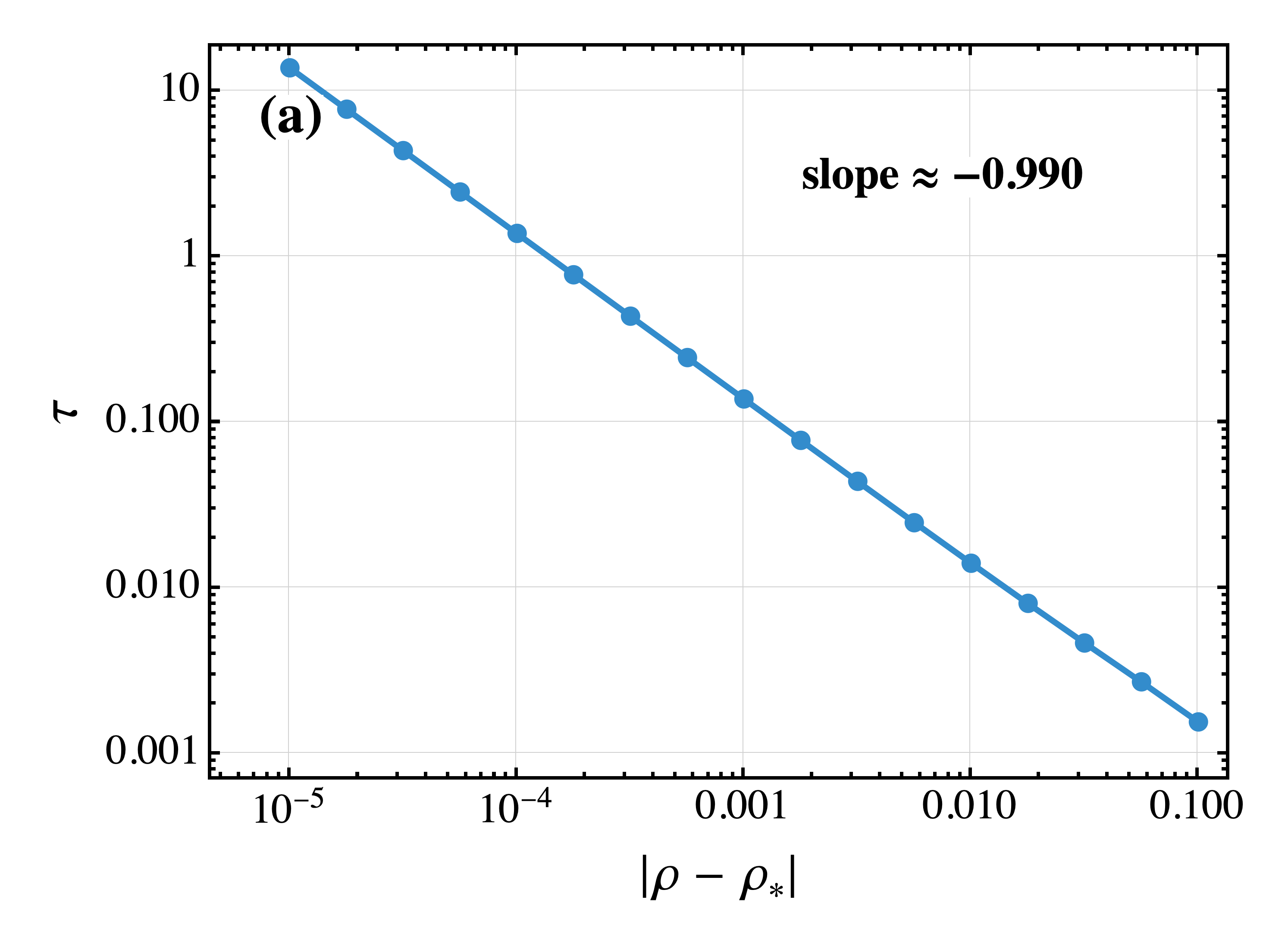} \\
			\includegraphics[scale=.5]{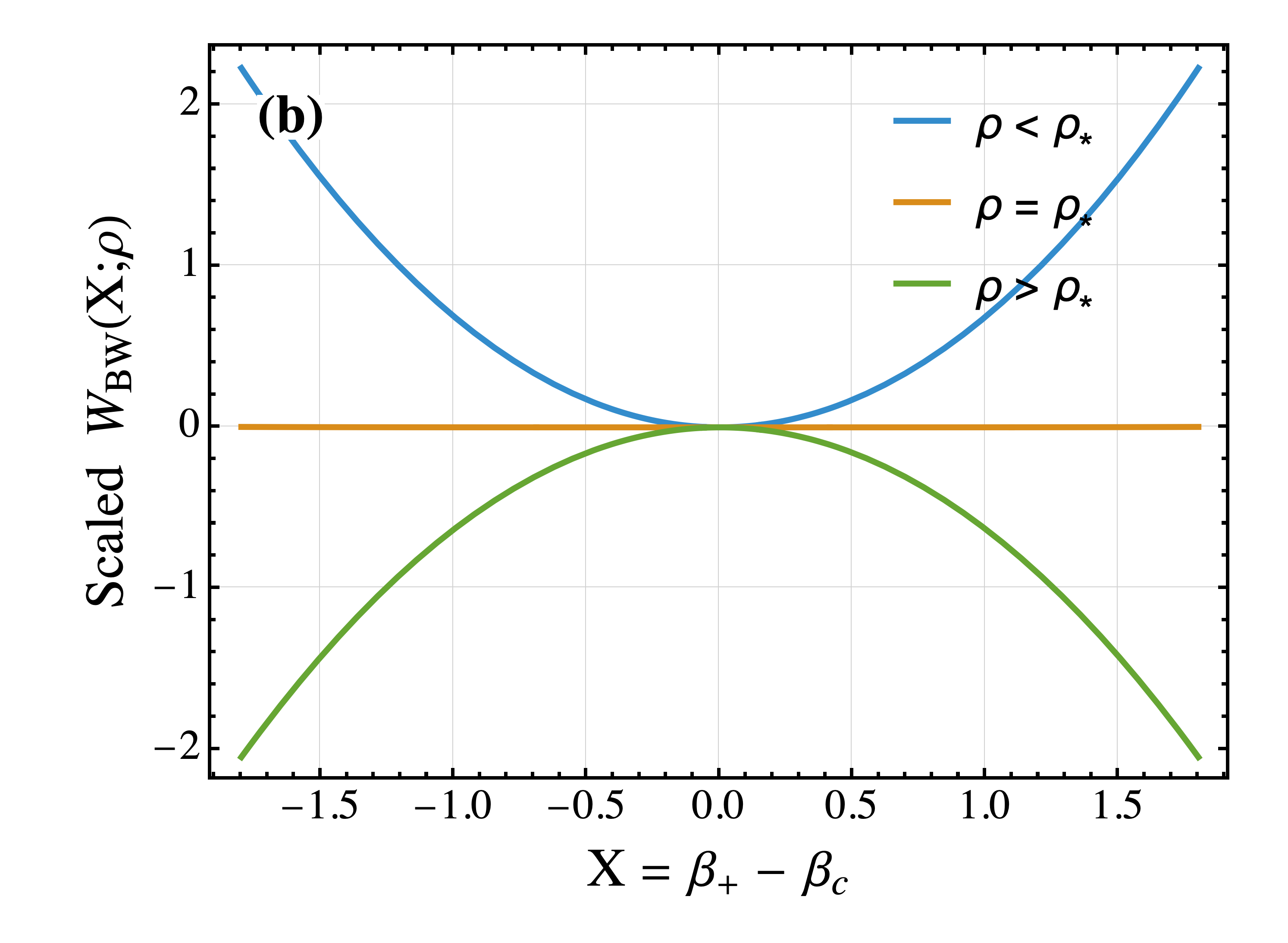} \\
		   \end{tabbing}

\caption{{\it \footnotesize  
Off-shell thermal structure and critical slowing down of the inter-horizon mode.
{\bf (a)} Critical slowing down of the thermal mode, displaying the scaling \(\tau \sim |\rho-\rho_*|^{-1}\) with fitted slope close to \(-1\), in agreement with the analytic prediction.
{\bf (b)} Bragg--Williams functional \(W_{\rm BW}(X;\rho)\) for representative values \(\rho<\rho_*\), \(\rho=\rho_*\), and \(\rho>\rho_*\), showing the change of local convexity at the critical ratio and the flattening of the thermal landscape at \(\rho=\rho_*\).
}  }\label{fig2}
\end{center}
\end{figure}

The critical slowing down of the inter-horizon thermal mode is shown in Fig.\ref{fig2}(a), whose fitted slope confirms the scaling \(\tau\sim |\rho-\rho_*|^{-1}\).

\section{Off-Shell  Thermal Description}

To encode the critical thermal structure off shell, we introduce a minimal effective Bragg--Williams functional for the affinity \(X\). This construction is not derived microscopically from Einstein--Maxwell dynamics; rather, it is the simplest off-shell completion compatible with the local relaxation law obtained above.  
Indeed, a Bragg--Williams functional for the thermal affinity \cite{Banerjee2010} reads
\begin{equation}
W_{\rm BW}(X;\rho)
=
W_0(\rho)
+\frac{1}{2}\mathcal K_L(\rho)X^2
+\frac{\lambda}{4}X^4
+O(X^6),
\qquad
\lambda>0.
\label{eq:WBW}
\end{equation}
Here the quartic term is the minimal regularizing off-shell completion compatible with the even truncation in the thermal mode \(X\). 
 The corresponding off-shell thermal landscape is displayed in Fig.\ref{fig2}(b), where the change of local convexity across the critical ratio is clearly visible.

The curvature at the origin is
\begin{equation}
\frac{\partial^2 W_{\rm BW}}{\partial X^2}\bigg|_{X=0}
=
\mathcal K_L(\rho),
\label{eq:WBWcurvature}
\end{equation}
so the same coefficient that controls the linear relaxation also governs the local convexity of the off-shell thermal landscape. The critical ratio \(\rho_*\) is thus the point at which the Bragg--Williams landscape becomes locally flat at the lukewarm point.

The corresponding Landau--Khalatnikov dynamics is
\begin{equation}
\dot X=-L\,\frac{\partial W_{\rm BW}}{\partial X},
\label{eq:LKdyn}
\end{equation}
which reproduces Eq.~\eqref{eq:mainmode} at linear order. If \(\mathcal K_L<0\), the truncated off-shell landscape develops the formal extrema
\begin{equation}
X_\pm=\pm\sqrt{-\frac{\mathcal K_L(\rho)}{\lambda}},
\label{eq:Xpm}
\end{equation}
which should be interpreted only as off-shell extrema of the effective thermal landscape, not as exact stationary RN--dS solutions.

We then elevate this deterministic effective dynamics to a trajectory-level description through an Onsager--Machlup action, understood here as an effective stochastic extension of the thermal mode rather than as a microscopic fluctuation theory of the full spacetime geometry.
  Consider the additive-noise process \cite{OnsagerMachlup1953a,OnsagerMachlup1953b}
\begin{equation}
\dot X
=
-L\,\frac{\partial W_{\rm BW}}{\partial X}
+\eta(t),
\qquad
\langle \eta(t)\eta(t')\rangle = 2D\,\delta(t-t'),
\label{eq:langevin}
\end{equation}
where \(D\) is an effective noise amplitude. We do not assume a microscopic fluctuation--dissipation relation here. Up to the standard normalization and boundary terms appropriate to the additive-noise Onsager--Machlup construction, the probability of a trajectory \(X(t)\) is controlled by
\begin{equation}
S_{\rm OM}[X]
=
\frac{1}{4D}
\int dt\,
\Big[
\dot X
+
L\big(\mathcal K_L(\rho)X+\lambda X^3\big)
\Big]^2.
\label{eq:SOM}
\end{equation}
Near the lukewarm manifold, the linearized action is
\begin{equation}
S_{\rm OM}^{\rm lin}[X]
=
\frac{1}{4D}
\int dt\,
\Big[
\dot X
+
L\mathcal K_L(\rho)X
\Big]^2.
\label{eq:SOMlin}
\end{equation}
Its most probable trajectory reproduces Eq.~\eqref{eq:mainmode}. At the critical ratio \(\rho=\rho_*\), the drift vanishes and the action reduces to
\begin{equation}
S_{\rm OM}^{\rm crit}[X]
=
\frac{1}{4D}
\int dt\,\dot X^2.
\label{eq:SOMcrit}
\end{equation}
The inter-horizon thermal mode thus becomes marginal at the critical point: the restoring drift disappears, and the effective trajectory weight becomes purely kinetic at linear order.

Two remarks are important. First, the present construction is explicitly restricted to the fixed-charge sector. Indeed, on the lukewarm manifold,
\begin{equation}
\Phi_+=\frac{Q}{r_+}=\frac{r_c}{r_++r_c},
\qquad
\Phi_c=\frac{Q}{r_c}=\frac{r_+}{r_++r_c},
\label{eq:potentials}
\end{equation}
so that \(\Phi_+\neq \Phi_c\) whenever \(r_+\neq r_c\). The lukewarm condition therefore defines a thermal equilibrium between the two horizons, but not a full thermo-electric equilibrium. Second, the Bragg--Williams and Onsager--Machlup structures introduced above are effective descriptions of the inter-horizon thermal mode in state space, not claims about the explicit spacetime dynamics of a fluctuating time-dependent black hole geometry.
\section{Conclusion}
We conclude that the lukewarm RN--dS family supports a sharply defined nonequilibrium thermal structure. The exact branch
\begin{equation}
M=Q=\frac{r_+r_c}{r_++r_c},
\qquad
\Lambda=\frac{3}{(r_++r_c)^2},
\label{eq:exactbranchfinal}
\end{equation}
provides the geometric anchor, but the deeper physical content lies in the effective two-horizon thermal dynamics it supports. In the fixed-charge sector, the lukewarm family is the exact zero-dissipation thermal manifold of the state-space evolution. 
This manifold carries a distinguished inter-horizon thermal mode whose stability changes at the exact ratio $\rho_*$. At this point, the relaxation time diverges as
\begin{equation}
\tau \sim |\rho-\rho_*|^{-1}.
\end{equation}
The associated Bragg--Williams landscape and Onsager--Machlup action show that the lukewarm manifold is not merely a special equal-temperature locus, but a critical thermal manifold endowed with a soft nonequilibrium mode. This perspective opens a route toward a broader nonequilibrium thermodynamics of multi-horizon black hole spacetimes and, potentially, toward topological reinterpretations of thermal stability structures \cite{WeiLiu2022,AliElMoumniKhalloufiMasmar2024}.

\begin{acknowledgments}
H. El Moumni acknowledges the networking support of COST Action CA22113 (Fundamental challenges in theoretical physics), CA21136 (Addressing observational tensions in cosmology with systematics and fundamental physics), and CA23130 (Bridging high and low energies in search of quantum gravity). He also thanks the Institute of Physics for its support. This work was carried out under the project UIZ 2025 Scientific Research Projects: PRJ-2025-81.
\end{acknowledgments}
\bibliography{refs}

\end{document}